\documentclass[conference]{IEEEtran}
\usepackage{amsmath}

\usepackage{amsmath,amssymb,epsfig,psfrag,cite,subfigure}
\include{macros}
\usepackage{graphicx}
\usepackage{epstopdf}

\usepackage{bm}

\usepackage{geometry}
\usepackage{tcolorbox}

\usepackage{accents}

\allowdisplaybreaks

\usepackage{soul}

\usepackage{amsthm}

\usepackage{nicefrac}

\usepackage{lipsum,multicol}

\usepackage{algorithm}
\usepackage{algpseudocode}

\algdef{SE}[SUBALG]{Indent}{EndIndent}{}{\algorithmicend\ }%
\algtext*{Indent}
\algtext*{EndIndent}

\usepackage{enumitem}
\usepackage{mwe}    
\usepackage{epsfig,psfrag}
\usepackage{subfigure}
\usepackage{color}
\usepackage{url}
\usepackage{mathtools,xparse}

\usepackage{gensymb}

\usepackage[multiple]{footmisc}

\usepackage{xcolor}

\theoremstyle{remark}

\newtheoremstyle{mytheoremstyle} 
    {\topsep}                    
    {\topsep}                    
    {\upshape}                   
    {.5em}                           
    {\itshape}                   
    {.}                          
    {.5em}                       
    {}  

\theoremstyle{plain}

\newtheoremstyle{iremark}
  {\topsep}   
  {\topsep}   
  {\upshape}  
  {0.2in}       
  {\itshape}  
  {.}         
  {5pt plus 1pt minus 1pt} 
  {\thmname{#1}\thmnumber{ \itshape#2}\thmnote{ (#3)}} 

\theoremstyle{definition}
\newtheorem*{proof}{Proof}

\DeclareFontFamily{U}{mathx}{\hyphenchar\font45}
\DeclareFontShape{U}{mathx}{m}{n}{
	<5> <6> <7> <8> <9> <10>
	<10.95> <12> <14.4> <17.28> <20.74> <24.88>
	mathx10
}{}
\DeclareSymbolFont{mathx}{U}{mathx}{m}{n}
\DeclareFontSubstitution{U}{mathx}{m}{n}

\DeclarePairedDelimiter\abs{\lvert}{\rvert}%

\makeatletter \renewcommand\d[1]{\ensuremath{%
		\;\mathrm{d}#1\@ifnextchar\d{\!}{}}}
\makeatother

\makeatletter
\newcommand*\rel@kern[1]{\kern#1\dimexpr\macc@kerna}
\newcommand*\widebar[1]{%
  \begingroup
  \def\mathaccent##1##2{%
    \rel@kern{0.8}%
    \overline{\rel@kern{-0.8}\macc@nucleus\rel@kern{0.2}}%
    \rel@kern{-0.2}%
  }%
  \macc@depth\@ne
  \let\math@bgroup\@empty \let\math@egroup\macc@set@skewchar
  \mathsurround\z@ \frozen@everymath{\mathgroup\macc@group\relax}%
  \macc@set@skewchar\relax
  \let\mathaccentV\macc@nested@a
  \macc@nested@a\relax111{#1}%
  \endgroup
}
\makeatother

\usepackage{scalerel,stackengine}
\stackMath
\newcommand\widecheck[1]{%
\savestack{\tmpbox}{\stretchto{%
  \scaleto{%
    \scalerel*[\widthof{\ensuremath{#1}}]{\kern-.6pt\bigwedge\kern-.6pt}%
    {\rule[-\textheight/2]{1ex}{\textheight}}
  }{\textheight}%
}{0.5ex}}%
\stackon[1pt]{#1}{\scalebox{-1}{\tmpbox}}%
}

\newcommand{\norm}[1]{\left\lVert#1\right\rVert}

\newcommand{\thn}[1]{ {#1^{\rm{th} } } }

\newcommand{\Eee}{\mathbb{E}}

\newcommand{\snr}{{\rm{SNR}}}

\newcommand{\rxii}{R_{\xi \xi}}
\newcommand{\deltat}{\Delta t}

\newcommand{\sigmaxi}{\sigma^2_{\xi}}

\newcommand{\Tsym}{ T_{\rm{sym}} }
\newcommand{\Ts}{ T_{\rm{s}} }
\newcommand{\Tcp}{ T_{\rm{cp}} }

\newcommand{\FF}{ \mathbf{F} }

\newcommand{\deltaf}{ \Delta f }

\newcommand{\fc}{ f_c }

\newcommand{\xnm}{ x_{n,m} }

\newcommand{\stilde}{ \widetilde{s} }
\newcommand{\ytilde}{ \widetilde{y} }

\newcommand{\mtN}{{\mathcal{N}}}

\newcommand{\fdb}{ f_{\rm{3dB}} }

\newcommand{\floop}{ f_{\rm{loop}} }

\newcommand{\taubar}{  \widebar{\tau} }
\newcommand{\nubar}{  \widebar{\nu} }
\newcommand{\mubt}{  \widetilde{\mub} }

\newcommand{\Dcal}{ \mathcal{D} }

\newcommand{\mub}{ \boldsymbol{\mu} }

\newcommand{\mtCN}{{\mathcal{CN}}}

\newcommand{\realp}[1]{ \Re \left\{#1\right\}  }
\newcommand{\imp}[1]{ \Im \left\{#1\right\}  }
\newcommand{\vecc}[1]{ {\rm{vec}}\left(#1\right)  }

\newcommand{\diag}[1]{ {\rm{diag}}\left(#1\right)  }

\newcommand{\Imatrix}{{ \boldsymbol{\mathrm{I}} }}

\newcommand{\cc}{ \mathbf{c} }
\newcommand{\bb}{ \mathbf{b} }

\newcommand{\boldzero}{{ {\boldsymbol{0}} }}

\newcommand{\JJ}{{ \mathbf{J} }}

\newcommand{\JJobs}{{ \JJ^{(\rm{o})} }}
\newcommand{\JJprior}{{ \JJ^{(\rm{p})} }}

\newcommand{\pt}{\widetilde{p}}

\newcommand{\detm}[1]{ {{{\rm{det}}\left( #1 \right)}}  }

\newcommand{\boldY}{ \mathbf{Y} }

\newcommand{\boldX}{ \mathbf{X} }

\newcommand{\boldW}{ \mathbf{W} }

\newcommand{\boldZ}{ \mathbf{Z} }

\newcommand{\boldA}{ \mathbf{A} }
\newcommand{\boldR}{ \mathbf{R} }

\newcommand{\boldB}{ \mathbf{B} }

\newcommand{\mcrb}{{\rm{MCRB}}}
\newcommand{\crb}{{\rm{CRB}}}
\newcommand{\lb}{{\rm{LB}}}
\newcommand{\lbavg}{{\rm{LB}}^{{\rm{avg}}}}

\newcommand{\qq}{ \mathbf{q} }
\newcommand{\yy}{ \mathbf{y} }

\newcommand{\zz}{ \mathbf{z} }

\newcommand{\boldXi}{ \mathbf{\Xi} }

\newcommand{\bxi}{ \boldsymbol{\xi} }

\newcommand{\transpose}[1]{ {#1}^{T} }

\newcommand{\complexset}[2]{ \mathbb{C}^{#1 \times #2}  }
\newcommand{\complexsett}{ \mathbb{C}  }
\newcommand{\realset}[2]{ \mathbb{R}^{#1 \times #2}  }

\newcommand{\alphabar}{ \widebar{\alpha} }

\newcommand{\alpharee}{ \alpha_{\rm{R}} }
\newcommand{\alphaime}{ \alpha_{\rm{I}} }

\newcommand{\alphabarre}{ \alphabar_{\rm{R}} }
\newcommand{\alphabarim}{ \alphabar_{\rm{I}} }

\newcommand{\etab}{ {\boldsymbol{\eta}} }

\newcommand{\etabbar}{ \widebar{\etab} }

\newcommand{\etabhat}{ \widehat{\boldsymbol{\eta}} }

\graphicspath{{RadarConf23/Figures_Conf/}}

\begin{document}
\bstctlcite{IEEEexample:BSTcontrol}

\title{On the Impact of Phase Noise on Monostatic Sensing in OFDM ISAC Systems}

\author{Musa Furkan Keskin\IEEEauthorrefmark{1}, Carina Marcus\IEEEauthorrefmark{2}, Olof Eriksson\IEEEauthorrefmark{2}, Henk Wymeersch\IEEEauthorrefmark{1}, and Visa Koivunen\IEEEauthorrefmark{3}\\
\IEEEauthorrefmark{1}Chalmers University of Technology, Sweden, \IEEEauthorrefmark{2}Veoneer Sweden AB, Sweden,
\IEEEauthorrefmark{3}Aalto University, Finland}



\maketitle

\begin{abstract}
     Phase noise (PN) can become a major bottleneck for integrated sensing and communications (ISAC) systems towards 6G wireless networks. In this paper, we consider an OFDM ISAC system with oscillator imperfections and investigate the impact of PN on monostatic sensing performance by performing a misspecified Cram\'{e}r-Rao bound (MCRB) analysis. Simulations are carried out under a wide variety of operating conditions with regard to SNR, oscillator type (free-running oscillators (FROs) and phase-locked loops (PLLs)), 3-dB bandwidth of the oscillator spectrum, PLL loop bandwidth and target range. The results provide valuable insights on when PN leads to a significant degradation in range and/or velocity accuracy, establishing important guidelines for hardware and algorithm design in 6G ISAC systems.

	\textit{Index Terms--} OFDM, integrated sensing and communications, phase noise, misspecified Cram\'{e}r-Rao bound.
\end{abstract}

\section{Introduction}\label{sec_intro}
\vspace{-0.05in}


Integrated sensing and communications (ISAC) is expected to be one of the enablers for 6G communication systems, both to support existing communication functions and to enable novel sensing applications \cite{Fan_ISAC_6G_JSAC_2022}. Dual-functioning monostatic systems are of particular interest, e.g., in automotive and IoT applications, due to their tight integration, low cost, and efficient spectrum usage \cite{JCAS_Survey_2022}. On the other hand, since such systems operate at mmWave frequencies (30--100 GHz), they are sensitive to hardware impairments, such as phase noise (PN), carrier frequency offsets, power amplifier nonlinearity \cite{hexax_pimrc_2021,RF_JCS_2021}, which can severely degrade the sensing performance, both in terms of detection and tracking.

Among the various hardware impairments, PN stands out, since it is time-varying and, in monostatic sensing, has target-dependent statistics \cite{ofdm_pn_2022}. The importance of PN is underlined by the intense research activities in radar-centric systems, such as FMCW-based radars \cite{PN_FMCW_2019,Canan_SPM_2020}. For communication-centric ISAC systems, PN is well understood only from the communication receiver perspective, where the PN statistics are independent of the objects in the propagation channel, and standard methods (either pilot-based or blind) can be utilized to mitigate the PN effect \cite{PN_OFDM_Sayed_TSP_2007,VI_PN_TSP_2007,PN_Spectral_ICI_TSP_2010,PN_OFDM_TSP_2017}. In sharp contrast, the impact of PN on OFDM ISAC sensing remains relatively underexplored \cite{ofdm_pn_2022,OFDM_JRC_PN_JLT_2022,SC_OFDM_PN}. 

In this paper, we carry out a detailed investigation of monostatic sensing performance in OFDM ISAC systems under the impact of PN by leveraging tools from estimation theory under model misspecification \cite{Richmond_TSP_2015,Fortunati2017}. Our specific contributions are as follows:
\begin{itemize}
    \item We derive the misspecified Cram\'{e}r-Rao bound (MCRB) and the corresponding lower-bound (LB) \cite{Fortunati2017} on range and velocity estimation for OFDM radar with PN, considering mismatch between a true model with PN and an assumed model without PN.
    \item Under correct specification of the observation model (i.e., when the receiver is aware of the existence of PN), we derive the hybrid CRB \cite{hybrid_CRB_LSP_2008,Hybrid_ML_MAP_TSP} on range, velocity and PN estimation.
    \item To quantify PN-induced accuracy degradations through a comparative analysis of the LB and the hybrid CRB, we conduct extensive simulations under a broad range of operating conditions pertaining to SNR, oscillator type/parameters and target range, which offer valuable insights into scenarios with significant impact of PN on radar performance.
\end{itemize}


\section{System Model}\label{sec_sys_mod}

\begin{figure}
    \centering
    \includegraphics[width=1\columnwidth]{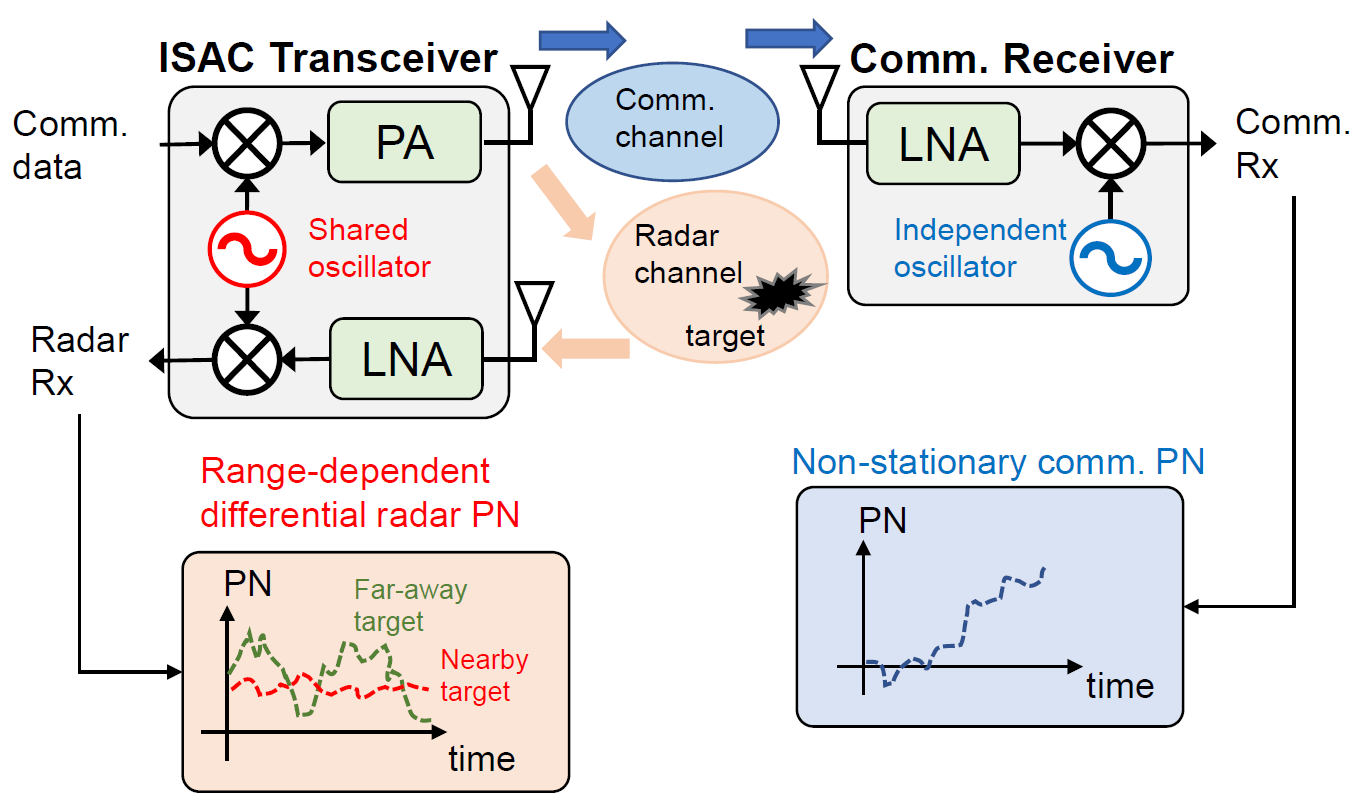}
    \caption{OFDM ISAC system with an ISAC transceiver and a communications receiver. Self-correlated (differential) PN process in the radar receiver due to the use of the shared oscillator in the ISAC transceiver leads to the range correlation effect (i.e., range-dependent PN statistics), unlike the range-independent non-stationary PN process in the communications receiver.}
    \label{fig_scenario}
    \vspace{-0.2in}
\end{figure}

We consider an OFDM ISAC system consisting of an ISAC transceiver and a communications receiver (RX), as shown in Fig.~\ref{fig_scenario}. The ISAC transceiver includes \textit{(i)} a conventional OFDM transmitter (TX) to send data symbols to the communications RX, and \textit{(ii)} a radar RX on the joint hardware platform to process the backscattered signals for target detection and parameter estimation \cite{RadCom_Proc_IEEE_2011}. The sensing configuration is monostatic since the TX and the radar RX are co-located. In addition, the communications RX performs standard OFDM receive operations, such as channel estimation, frequency synchronization and data detection \cite{PN_mmWave_OFDM_TWC_2022}. In the joint ISAC transceiver hardware, the TX and radar RX shares the same oscillator, which is assumed to be imperfect and impaired by phase noise (PN) \cite{PN_OFDM_PLL_TCOM_2007,OFDM_PN_HCRB_2014}. In this section, we derive OFDM transmit and radar receive signal models in the presence of PN and provide a statistical characterization of PN in the radar receiver.

\subsection{Transmit Signal Model}\label{sec_transmit}
Consider an OFDM frame with $M$ symbols, $N$ subcarriers, total symbol duration $\Tsym = \Tcp + T$, elementary symbol duration $T$, cyclic prefix (CP) duration $\Tcp$ and subcarrier spacing $\deltaf = 1/T$. The complex baseband OFDM transmit signal can be expressed as \cite{RadCom_Proc_IEEE_2011,General_Multicarrier_Radar_TSP_2016}
\begin{align}
    s(t) = \frac{1}{\sqrt{N}} \sum_{m=0}^{M-1}  \sum_{n = 0}^{N-1}  \xnm \, e^{j 2 \pi n \deltaf t} \Pi\left(\frac{t - m\Tsym}{\Tsym}\right) ~,
\end{align}
where $\xnm \in \complexsett$ is the data symbol on the $\thn{n}$ subcarrier of the $\thn{m}$ symbol and $\Pi(t)$ is a rectangular pulse assuming the value $1$ for $t \in \left[0, 1 \right]$ and $0$ otherwise. Let $\phi(t)$ denote the PN process in the oscillator and $\fc$ the carrier frequency. Then, the upconverted transmit signal is given by \cite{PN_2006}
\begin{equation}\label{eq_passband_st}
\stilde(t) = \Re \left\{  s(t) e^{j \left[ 2 \pi \fc t + \phi(t)\right]} \right\} ~.
\end{equation}

\subsection{Radar Receive Signal Model}\label{sec_radar_rec}
We consider a single target with round-trip delay $\tau = 2 R/c$, normalized Doppler shift $\nu = 2 v/c$ and complex channel gain $\alpha$, where $R$, $v$ and $c$ represent the distance, radial velocity and speed of propagation, respectively. For the transmit signal in \eqref{eq_passband_st}, the passband signal at the radar RX is given by
\begin{align}\label{eq_rec_passband}
    \ytilde(t) = \Re \left\{ \alpha \, s(t-\tau(t)) e^{j \left[ 2 \pi \fc (t-\tau(t)) + \phi(t-\tau(t))\right]} \right\} ~,
\end{align}
where the Doppler shift causes a time-varying delay $\tau(t) = \tau - \nu t$. The downconversion of \eqref{eq_rec_passband} occurs through the noisy oscillator, which is equivalent to multiplication by $e^{-j (2 \pi \fc t + \phi(t))}$ \cite{PN_SI_TSP_2017}, leading to the equivalent complex baseband signal \cite{SPM_PN_2019}
\begin{align}\label{eq_rec_baseband}
    y(t) &= \alpha \, s(t-\tau(t)) e^{-j 2 \pi \fc \tau} e^{j 2 \pi \fc \nu t }  e^{j \left[ \phi(t-\tau(t)) - \phi(t)\right] } ~.
\end{align}
Assuming $ \lvert \nu \rvert \ll 1/N$ \cite{Firat_OFDM_2012,ICI_OFDM_TSP_2020,MIMO_OFDM_ICI_JSTSP_2021} (which holds, e.g., for typical vehicular ISAC scenarios \cite{ofdm_pn_2022}) and small time-bandwidth product \cite{OFDM_ICI_TVT_2017}, the signal in \eqref{eq_rec_baseband} becomes
\begin{align} \label{eq_rec_baseband2}
    y(t) &= \alpha \, s(t-\tau) e^{-j 2 \pi \fc \tau} e^{j 2 \pi \fc \nu t }  w(t, \tau) ~,
\end{align}
where $w(t, \tau)$ is the multiplicative PN process, i.e.,
\begin{align} \label{eq_pn_def}
    w(t, \tau) \triangleq e^{j \left[ \phi(t-\tau) - \phi(t)\right] } ~.
\end{align}

\subsection{Fast-Time/Slow-Time Representation with Phase Noise}\label{sec_fast_slow}
Following traditional OFDM radar processing for \eqref{eq_rec_baseband2} (i.e., removing the CP in the $\thn{m}$ symbol and sampling at $t = m\Tsym + \Tcp + \ell T / N$ for $\ell = 0, \ldots, N-1$), and adopting the standard assumptions $\Tcp \geq \tau$ \cite{Firat_OFDM_2012,OFDM_Radar_Phd_2014,SPM_JRC_2019} and $\fc T \nu \ll 1$ \cite{Passive_OFDM_2010,OTFS_RadCom_TWC_2020,OFDM_DFRC_TSP_2021}, the received signal for the $\thn{m}$ symbol can be written as \cite{ICI_OFDM_TSP_2020,MIMO_OFDM_ICI_JSTSP_2021,ofdm_pn_2022}
\begin{align}\label{eq_rec_bb2}
    y_{\ell,m} &= \alpha  \, e^{j 2 \pi \fc m \Tsym \nu  }  w_{\ell,m}(\tau) \\ \nonumber &~~\times \frac{1}{\sqrt{N}}  \sum_{n = 0}^{N-1}  \xnm \, e^{j 2 \pi n \frac{\ell}{N}} e^{-j 2 \pi n \deltaf \tau} ~,
\end{align}
where the sampled version of the PN term $w(t, \tau)$ in \eqref{eq_rec_baseband2} for $t = m\Tsym + \Tcp + \ell T / N$ is represented by $w_{\ell,m}(\tau)$. Let us define the frequency-domain and temporal (slow-time) steering vectors as
\begin{align} \label{eq_steer_delay}
	\bb(\tau) & \triangleq  \transpose{ \left[ 1, e^{-j 2 \pi \deltaf \tau}, \ldots,  e^{-j 2 \pi (N-1) \deltaf  \tau} \right] } ~, \\ \label{eq_steer_doppler}
	\cc(\nu) & \triangleq \transpose{ \left[ 1, e^{-j 2 \pi f_c \Tsym \nu }, \ldots,  e^{-j 2 \pi f_c (M-1) \Tsym \nu } \right] } ~.
\end{align}
Then, we can organize the observations in \eqref{eq_rec_bb2} 
into a fast-time/slow-time observation matrix \cite{MIMO_OFDM_ICI_JSTSP_2021}
\begin{align} \label{eq_ym_all_multi}
    \boldY = \alpha \, \boldW \odot \FF_N^{H} \Big(\boldX \odot \bb(\tau) \cc^{H}(\nu) \Big)  + \boldZ ~,
\end{align}
where $\boldW \in \complexset{N}{M}$ with $\left[ \boldW \right]_{\ell,m} \triangleq w_{\ell,m}(\tau)$ is the multiplicative PN matrix, $\FF_N \in \complexset{N}{N}$ is the unitary DFT matrix with $\left[ \FF_N \right]_{\ell,n} = \frac{1}{\sqrt{N}} e^{- j 2 \pi n \frac{\ell}{N}} $, $\boldX \in \complexset{N}{M}$ contains the data symbols with $\left[ \boldX \right]_{n,m} \triangleq \xnm$, $\boldY \in \complexset{N}{M}$ with $\left[ \boldY \right]_{\ell,m} \triangleq y_{\ell,m} $, and $\boldZ \in \complexset{N}{M}$ is additive complex white Gaussian sensor noise with $\vecc{\boldZ} \sim \mtCN(\boldzero, \allowbreak 2\sigma^2 \Imatrix ) $. 


\subsection{Phase Noise Statistics}\label{sec_pn_statistics}
Let us define the differential/self-correlated PN process \cite{Demir_PN_2006,DPN_93,PN_2006} in \eqref{eq_pn_def} as
\begin{equation}\label{eq_dpn}
\xi(t, \tau) \triangleq \phi(t) - \phi(t-\tau) ~,
\end{equation} 
which is a stationary process with statistics that depend on the delay $\tau$ \cite[Sec.~IV]{Demir_PN_2006}, \cite{PN_2006,ofdm_pn_2022}:
\begin{equation}\label{eq_dpn_stat}
\xi(t, \tau) \sim \mtN(0, \sigmaxi(\tau))~.
\end{equation}
In \eqref{eq_dpn_stat}, $\sigmaxi(\tau)$ represents the delay-dependent variance of $\xi(t, \tau) $ whose form depends on the oscillator type, i.e., free-running oscillator (FRO) or phase-locked loop (PLL). In particular, we have \cite{Demir_PN_2006,PN_2006,PN_OFDM_PLL_TCOM_2007}
\begin{align} \label{eq_fro_pll_variance}
    \sigmaxi(\tau) = \begin{cases} 4 \pi \fdb \abs{\tau}~, & \text{FRO} 
    \\
    \frac{2\fdb}{\floop} \left(1 - e^{-2 \pi \floop \abs{\tau}} \right)~, & \text{PLL}
    \end{cases} ~,
\end{align}
where $\fdb$ and $\floop$ denote the $3 \, \rm{dB}$ bandwidth of the Lorentzian oscillator spectrum and the loop bandwidth of PLL, respectively.

We denote by $\bxi \in \realset{NM}{1}$ the discrete-time version of the PN process in \eqref{eq_dpn} over $NM$ time-domain samples in the considered OFDM frame. According to \eqref{eq_pn_def} and \eqref{eq_ym_all_multi}, we have the relation $\vecc{\boldW} = e^{-j \bxi}$. Based on \cite{ofdm_pn_2022}, the statistics of $\bxi$ are given by
\begin{align}\label{eq_bxi_stat}
    \bxi \sim \mtN(\boldzero, \boldR(\tau))~,
\end{align}
where $\boldR(\tau) \in \realset{NM}{NM}$ is the delay-dependent covariance matrix of $\bxi$, with the entries
\begin{align}\label{eq_rtau_entries}
    \left[ \boldR(\tau) \right]_{i_1,i_2} = \rxii(\deltat_{i_1 i_2}, \tau) ~.
\end{align}
In \eqref{eq_rtau_entries},
\begin{align}\label{eq_exp_xi2}
     \rxii(\deltat, \tau) &= \frac{ \sigmaxi(\tau+\deltat) + \sigmaxi(\tau-\deltat) }{2} - \sigmaxi(\deltat) ~,
\end{align}
is the correlation function of $\xi(t, \tau)$ \cite[Lemma~1]{ofdm_pn_2022} and $\deltat_{i_1 i_2} \triangleq (i_1-i_2) \Ts +(m_1-m_2)\Tcp$ for $(i_1, i_2) = (n_1 + m_1 N, n_2 + m_2 N)$, where $0 \leq n_1, n_2 \leq N-1$ and $0 \leq m_1, m_2 \leq M-1$ are the fast-time and slow-time sample indices, respectively, and $\Ts = T/N$ is the sampling interval.


\subsection{Goals}\label{sec_formulation}
Given the observation model in \eqref{eq_ym_all_multi} and the PN statistics in \eqref{eq_bxi_stat}, our goal is to characterize the performance of range and velocity estimation in the presence of PN under a wide variety of operating conditions with regard to SNR, target range and oscillator quality (i.e., $\fdb$ and $\floop$). To accomplish this goal, we resort to standard CRB and MCRB as analytical tools.

\section{CRB Analysis under Phase Noise}\label{sec_mcrb}
In this section, we derive the theoretical bounds on range and velocity estimation with and without PN, and under different levels of knowledge on the presence of PN and its statistics. 

\subsection{Deterministic CRB without Phase Noise}
In this part, we assume that the generative model in \eqref{eq_ym_all_multi} does not contain PN, i.e.,
\begin{align} \label{eq_ym_all_multi_pn_free}
    \boldY = \alpha \, \FF_N^{H} \Big(\boldX \odot \bb(\tau) \cc^{H}(\nu) \Big)  + \boldZ ~.
\end{align}
In vector form, \eqref{eq_ym_all_multi_pn_free} can be written as
\begin{align}\label{eq_yy_true_pn_free}
    \yy = \alpha \, \qq(\tau, \nu) + \zz ~,
\end{align}
where $\yy \triangleq \vecc{\boldY} \in \complexset{NM}{1}$, $\zz \triangleq \vecc{\boldZ} \in \complexset{NM}{1}$, and
\begin{align} 
\label{eq_qtaunu}
    \qq(\tau, \nu) &\triangleq \vecc{\FF_N^{H} \Big[\boldX \odot \bb(\tau) \cc^{H}(\nu) \Big] } \in \complexset{NM}{1} ~.
\end{align}
The unknown parameter vector in \eqref{eq_yy_true_pn_free}, which involves only deterministic parameters, is given by
\begin{align}\label{eq_eta_crb}
    \etab = \left[ \tau, \nu, \alpharee, \alphaime \right]^T \in \realset{4}{1} ~,
\end{align}
where $\alpharee \triangleq \realp{\alpha}$ and $\alphaime \triangleq \imp{\alpha}$. Then, the deterministic CRB on the variance of an unbiased estimator $\etabhat(\yy)$ of $\etab$ in \eqref{eq_eta_crb} can be expressed as \cite{kay1993fundamentals}
\begin{align} \label{eq_fim_det_crb}
    \Eee_{\yy}\left[ \left(\etabhat(\yy) - \etab \right) \left(\etabhat(\yy) - \etab \right)^T \right] \succeq \JJ^{-1} ~,
\end{align}
where $\JJ \in \realset{4}{4}$ is the Fisher information matrix (FIM), given by (using the Slepian-Bangs formula \cite[Eq.~(15.52)]{kay1993fundamentals} through the Gaussianity of the measurement in \eqref{eq_yy_true_pn_free})
\begin{align}\label{eq_fim}
    \left[\JJ \right]_{i,j} &=  \frac{1}{\sigma^2} \realp{ \frac{ \partial \mub^H(\etab) }{ \partial \eta_i } \frac{ \partial \mub(\etab) }{ \partial \eta_j }  }  ~,
\end{align}
where $\mub(\etab) \triangleq \alpha \, \qq(\tau, \nu)$ is the mean of $\yy$ in \eqref{eq_yy_true_pn_free} and $\eta_i \triangleq [\etab]_i$. Using the FIM in \eqref{eq_fim}, the CRBs on delay and Doppler estimation accuracy can be computed as
\begin{align} \label{eq_crb_delay_doppler_det}
    \crb_{\tau} = [\JJ^{-1}]_{1,1} ~, ~~    \crb_{\nu} = [\JJ^{-1}]_{2,2} ~.
\end{align}

\subsection{Hybrid CRB under Phase Noise}\label{sec_crb}
In this part, we consider the generative model in \eqref{eq_ym_all_multi} with PN, which can be expressed in vector form as
\begin{align}\label{eq_yy_vec}
    \yy = \alpha \, \boldXi \qq(\tau, \nu) + \zz ~,
\end{align}
where $\boldXi \triangleq \diag{e^{-j \bxi}} \in \complexset{NM}{NM}$. For the observation in \eqref{eq_yy_vec}, the unknown parameter vector is given by
\begin{align}\label{eq_eta_unk_doppler}
    \etab = \left[ \tau, \nu, \alpharee, \alphaime, \bxi  \right]^T \in \realset{(NM+4)}{1} ~,
\end{align}
which consists of both deterministic delay-Doppler-gain parameters and random PN. Then, the hybrid CRB on the variance of an unbiased estimator $\etabhat(\yy)$ of $\etab$ in \eqref{eq_eta_unk_doppler} can be written as \cite{hybrid_CRB_LSP_2008,Hybrid_ML_MAP_TSP}
\begin{align} \label{eq_hfim_}
    \Eee_{\yy,\bxi}\left[ \left(\etabhat(\yy) - \etab \right) \left(\etabhat(\yy) - \etab \right)^T \right] \succeq \JJ_{\rm{hyb}}^{-1} ~,
\end{align}
where $\JJ_{\rm{hyb}} \in \realset{(NM+4)}{(NM+4)}$ is the hybrid FIM (HFIM), given by
\begin{align}\label{eq_hfim}
    \JJ_{\rm{hyb}} = \JJobs + \JJprior ~,
\end{align}
with $\JJobs$ and $\JJprior$ representing the observation-related and the prior information-related HFIMs, respectively. Using
\cite[Eq.~(7)]{hybrid_CRB_LSP_2008} and the Slepian-Bangs formula \cite[Eq.~(15.52)]{kay1993fundamentals}, we obtain 
\begin{align} \label{eq_jo}
    \left[\JJobs \right]_{i,j} &= \Eee_{\bxi} \left[ \frac{1}{\sigma^2} \realp{ \frac{ \partial \mub^H(\etab) }{ \partial \eta_i } \frac{ \partial \mub(\etab) }{ \partial \eta_j }  } \right]  ~,
    \\ 
    \left[\JJprior \right]_{i,j} &= \Eee_{\bxi} \left[ \frac{ \partial \log p_{\bxi}(\bxi; \tau)  }{ \partial \eta_i } \frac{ \partial \log p_{\bxi}(\bxi; \tau)  }{ \partial \eta_j }  \right] ~,
\end{align}
where $\mub(\etab) \triangleq \alpha \, \boldXi \qq(\tau, \nu)$ is the mean of $\yy$ given $\bxi$ in \eqref{eq_yy_vec} and $p_{\bxi}$ is the pdf of PN as a function of delay, obtained from \eqref{eq_bxi_stat} as
\begin{align} \label{eq_pn_pdf}
    p_{\bxi}(\bxi; \tau) &= \frac{1}{\sqrt{(2 \pi)^{NM} \detm{\boldR(\tau)}  } } \exp\left\{ - \frac{ \bxi^T \boldR(\tau)^{-1} \bxi}{2 } \right\} ~.
\end{align}
Consequently, the hybrid CRBs on delay and Doppler estimation accuracy are given by
\begin{align} \label{eq_crb_delay_doppler}
    \crb_{\tau} = [\JJ_{\rm{hyb}}^{-1}]_{1,1} ~, ~~    \crb_{\nu} = [\JJ_{\rm{hyb}}^{-1}]_{2,2} ~.
\end{align}

\subsection{Misspecified CRB under Phase Noise}
The previous part considers a scenario in which the receiver is aware of the existence of PN in the observations and has the knowledge of its statistics in \eqref{eq_bxi_stat}. In this part, considering the same generative model as in Sec.~\ref{sec_crb}, we investigate a scenario where the receiver is unaware of the existence of PN (or, the probability model for PN is unknown) and applies standard processing suited to the ideal, PN-free model in \eqref{eq_yy_true_pn_free}. By leveraging the MCRB tool \cite{Fortunati2017, Fortunati2018Chapter4,MCRB_delay_ICASSP_2020}, we aim to quantify degradation in range/velocity accuracy under PN due to model misspecification between the \textit{true model with PN} and the \textit{assumed model without PN}.

\subsubsection{True and Assumed Models for Radar Observation}\label{sec_true_assumed}
In the MCRB nomenclature \cite{Fortunati2017}, the \textit{true model} corresponds to the one in \eqref{eq_yy_vec}, which involves the effect of PN, i.e.,
\begin{align}\label{eq_yy_true}
    \yy = \alphabar \, \boldXi \qq(\taubar, \nubar) + \zz ~,
\end{align}
where $\alphabar$, $\taubar$ and $\nubar$ represent the true values of the unknown parameters $\alpha$, $\tau$ and $\nu$, respectively.
The pdf of the true observation model in \eqref{eq_yy_true} for a given PN realization $\bxi$ is given by
\begin{align}\label{eq_p_true}
    p(\yy) = \frac{1}{(2\pi \sigma^2)^{NM} } \exp\left\{ - \frac{ \norm{\yy - \mub }^2 }{2 \sigma^2} \right\} ~,
\end{align}
where $\mub \triangleq \alphabar \, \boldXi \qq(\taubar, \nubar) \in \complexset{NM}{1}$.

In practice, PN is usually ignored in radar processing, leading to the \textit{assumed model} without the impact of PN:
\begin{align}\label{eq_yy_assumed}
    \yy = \alpha \, \qq(\tau, \nu) + \zz ~.
\end{align}
For the assumed model in \eqref{eq_yy_assumed}, the misspecified parametric pdf is given by \cite{Fortunati2017}
\begin{align}\label{eq_p_assumed}
    \pt(\yy  \lvert \etab) = \frac{1}{(2\pi \sigma^2)^{NM} } \exp\left\{ - \frac{ \norm{\yy - \mubt(\etab) }^2 }{2 \sigma^2} \right\} ~,
\end{align}
where $\mubt(\etab) \triangleq \alpha \, \qq(\tau, \nu) \in \complexset{NM}{1}$ and $\etab = \left[ \tau, \nu, \alpharee, \alphaime \right]^T$ is the unknown parameter vector.

\subsubsection{Pseudo-True Parameter}
To derive the expression for MCRB, we first define the \textit{pseudo-true parameter} \cite{Fortunati2017}
\begin{equation}\label{eq_eta0}
		\etab_0 = \arg \min_{\etab} ~ \Dcal \left( p(\yy) ~ \Vert ~ \pt(\yy \lvert \etab) \right),
\end{equation}
where $\Dcal \left( p(\yy) ~ \Vert ~ \pt(\yy \lvert \etab) \right)$ is the Kullback-Leibler (KL) divergence between the true pdf in \eqref{eq_p_true} and the assumed pdf in \eqref{eq_p_assumed}. Based on \cite[Lemma~1]{ozturk2022ris}, \eqref{eq_eta0} is equivalent to  
\begin{align}
		\etab_0 &= \arg \min_{\etab} ~ \norm{\mub - \mubt(\etab)}^2 ~,
		\\ \label{eq_eta02}
		(\alpha_0, \tau_0, \nu_0) &= \arg \min_{\alpha, \tau, \nu} ~ \norm{\mub - \alpha \, \qq(\tau, \nu)}^2 ~.
\end{align}
In \eqref{eq_eta02}, $\alpha_0$ is readily obtained as $\alpha_0 = \qq(\tau, \nu)^{\dagger} \mub$ as a function of $\tau$ and $\nu$, where $\boldA^{\dagger} = (\boldA^H \boldA)^{-1} \boldA^H$ denotes the pseudo-inverse of $\boldA$. Plugging $\alpha_0$ back into \eqref{eq_eta02}, we have
\begin{align}\nonumber
    (\tau_0, \nu_0) &= \arg \max_{\tau, \nu} ~ \frac{\abs{\mub^H \qq(\tau, \nu)}^2 }{\norm{\qq(\tau, \nu)}^2} = \arg \max_{\tau, \nu} ~ \abs{\mub^H \qq(\tau, \nu)}^2  ~,
\end{align}
where $\norm{\qq(\tau, \nu)}^2 = \norm{\boldX}_F^2$ is constant.

\subsubsection{MCRB and LB Derivation}
Using the pseudo-true parameter in \eqref{eq_eta0}, the MCRB matrix is given by \cite{Fortunati2017, Fortunati2018Chapter4}
\begin{align} \label{eq_mcrb_def}
		\mcrb(\etab_0) = \boldA_{\etab_0}^{-1} \boldB_{\etab_0} \boldA_{\etab_0}^{-1} \in \realset{4}{4} ~,
\end{align}
where the entries of $\boldA_{\etab_0} \in \realset{4}{4}$ and $\boldB_{\etab_0} \in \realset{4}{4}$ are given by
\begin{align}\label{eq_Aeta0}
		[\boldA_{\etab_0}]_{ij} &= \mathbb{E}_p\left\{\frac{\partial^2}{\partial \eta_i \partial \eta_j} \log \pt(\yy  \lvert \etab) \Big|_{\etab = \etab_0}\right\}, 
		\\ \label{eq_Beta0}
		[\boldB_{\etab_0}]_{ij} &= \mathbb{E}_p\left\{\frac{\partial}{\partial \eta_i } \log \pt(\yy  \lvert \etab) \frac{\partial}{\partial \eta_j } \log \pt(\yy  \lvert \etab) \Big|_{\etab =  \etab_0}\right\},
\end{align} 
with $\mathbb{E}_p\left\{\cdot\right\}$ denoting the expectation over the true pdf in \eqref{eq_p_true} (see \cite{ozturk2022ris} for derivation of $\boldA_{\etab_0}$ and $\boldB_{\etab_0}$). The MCRB provides a lower bound on the covariance matrix of any misspecified-unbiased (MS-unbiased) estimator of $\etab_0$, i.e., \cite{Fortunati2017}
\begin{align}
		\mathbb{E}_p\{(\etabhat(\yy)-\etab_0)(\etabhat(\yy)-\etab_0)^T\} \succeq  \mcrb(\etab_0),
\end{align}
where $\etabhat(\yy)$ is an MS-unbiased estimator of $\etab_0$, derived under the misspecified model, i.e., $\mathbb{E}_p\left\{ \etabhat(\yy) \right\} = \etab_0$. Based on the MCRB, the covariance matrix of any MS-unbiased estimator $\etabhat(\yy)$ with respect to the true parameter $\etabbar \triangleq [\alphabarre, \alphabarim, \taubar, \nubar]^T$ can be lower-bounded as \cite{Fortunati2017}
\begin{align}
		\mathbb{E}_p\{(\etabhat(\yy)-\etabbar)(\etabhat(\yy)-\etabbar)^T\} \succeq  \lb(\etab_0),
\end{align}
where \vspace{-0.2in}
\begin{align}\label{eq_lb}
	  \lb(\etab_0) = \mcrb(\etab_0)  + \overbrace{(\etabbar-\etab_0)(\etabbar-\etab_0)^T}^{\rm{Bias}}~.
\end{align}

\subsubsection{Expected Performance under PN}
The LB in \eqref{eq_lb} is a deterministic bound that can be used to characterize the performance of delay-Doppler estimation in the presence of PN for a given realization $\bxi$, where the observations are generated using the true model in \eqref{eq_p_true} and the estimator design is based on the assumed model in \eqref{eq_p_assumed}. We can obtain the expected performance by averaging the LB over the distribution of PN in \eqref{eq_bxi_stat} \cite{Richmond_TSP_2015}, i.e.,
\begin{align} \label{eq_lb_avg}
    \lbavg(\etab_0) = \mathbb{E}_{p_{\bxi}} \{ \lb(\etab_0) \} ~.
\end{align}
Using \eqref{eq_lb_avg}, the expected LBs on delay and Doppler estimation accuracy are given by
\begin{align}\label{eq_lb_delay_doppler}
    \lbavg_{\tau} = [\lbavg(\etab_0)]_{1,1} ~, ~~
    \lbavg_{\nu} = [\lbavg(\etab_0)]_{2,2} ~.
\end{align}


\section{Simulation Results}\label{sec_sim}
To evaluate the impact of PN on the radar performance, we consider the simulation setup in Table~\ref{tab_parameters} in compliance with 5G NR FR2 parameters \cite{TR_38211}. For the observations in \eqref{eq_ym_all_multi}, we define the SNR as $\snr = \abs{\alpha}^2/(2 \sigma^2)$ and the data symbols $\boldX$ are drawn randomly from the QPSK alphabet. In addition, the oscillator parameters are set to $\fdb = 100 \, \rm{kHz}$ and $\floop = 1 \, \rm{MHz}$ (in the case of PLL) \cite{Demir_PN_2006}. Moreover, a target with range $R = 50 \, \rm{m}$ and velocity $v = 20 \, \rm{m/s}$ is considered. For performance evaluation, we investigate the following theoretical bounds in our simulations:
\begin{itemize}
    \item \textbf{CRB (PN-free):} The deterministic CRBs on delay and Doppler estimation in \eqref{eq_crb_delay_doppler_det}.

    \item \textbf{CRB:} The hybrid CRBs on delay and Doppler estimation in \eqref{eq_crb_delay_doppler}.
    
    \item \textbf{LB:} The LB on delay and Doppler estimation in \eqref{eq_lb_delay_doppler}, computed using numerical averaging over $100$ PN realizations in \eqref{eq_lb_avg}.
    
\end{itemize}
The {\textbf{CRB (PN-free)}} serves as an ideal, hypothetical baseline to show performance losses due to PN with respect to the case where the receiver is aware/unaware of the existence of PN, corresponding to {\textbf{CRB}}/{\textbf{LB}}, respectively.

\begin{table}\footnotesize
\caption{OFDM Parameters}
\vspace{-0.05in}
\centering
    \begin{tabular}{|l|l|}
        \hline
        \textbf{Parameter} & \textbf{Value} \\ \hline
        Carrier Frequency, $\fc$  & $28 \, \rm{GHz}$ \\ \hline
        Total Bandwidth, $B$ & $30.72 \, \rm{MHz}$ \\ \hline
        Number of Subcarriers, $N$ & $256$ \\ \hline
        Number of Symbols, $M$ & $10$ \\ \hline
        Subcarrier Spacing, $\deltaf$ & $120 \, \rm{kHz}$  \\ \hline
        Symbol Duration, $T$ & $8.33 \, \mu \rm{s}$  \\ \hline
        Cyclic Prefix Duration, $\Tcp$ & $0.58 \, \mu \rm{s}$  \\ \hline
        Total Symbol Duration, $\Tsym$ & $8.91 \, \rm{\mu s}$  \\ \hline
    \end{tabular}
    \label{tab_parameters}
    \vspace{-0.1in}
\end{table}

\subsection{Performance with respect to SNR}
\begin{figure}[t]
        \begin{center}
        \subfigure[]{
			 \label{fig_range_rmse_r30_v20_f3dB_200_FRO}
			 \includegraphics[width=0.38\textwidth]{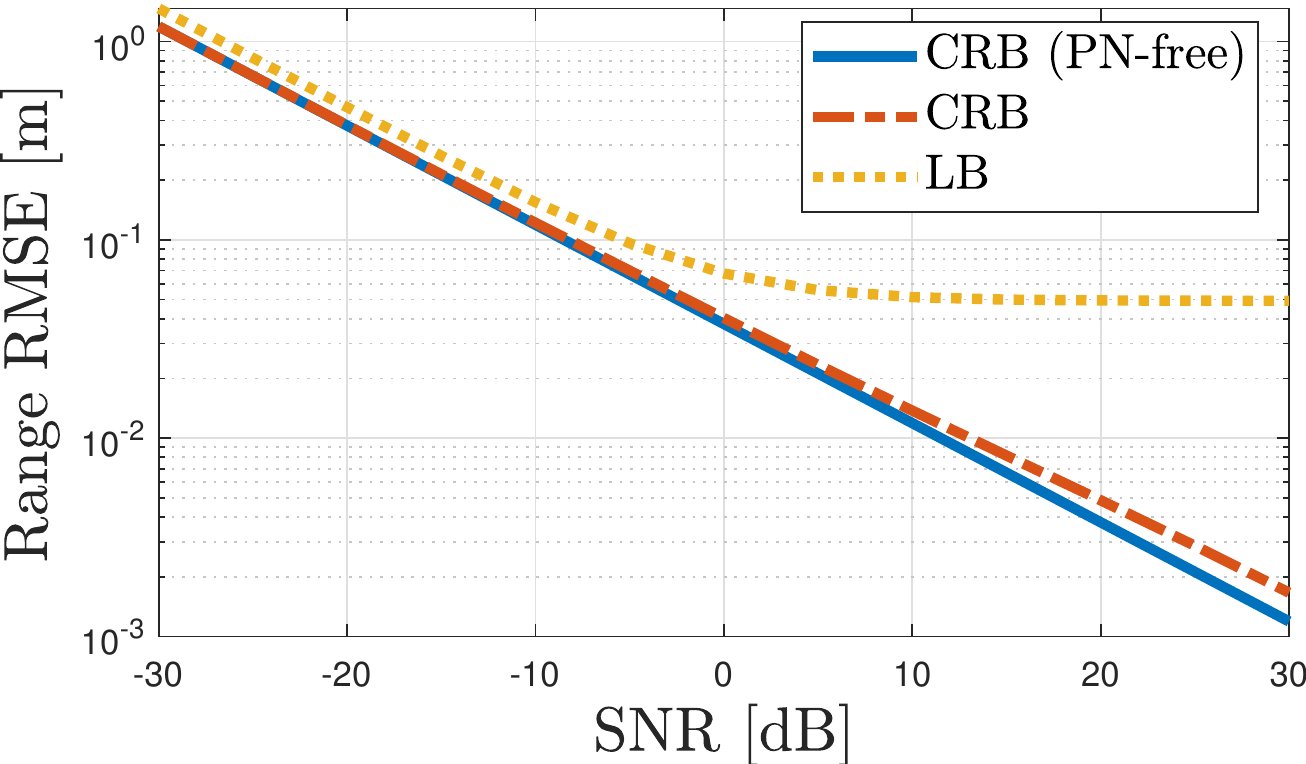}
		}
        \subfigure[]{
			 \label{fig_vel_rmse_r30_v20_f3dB_200_FRO}
			 \includegraphics[width=0.38\textwidth]{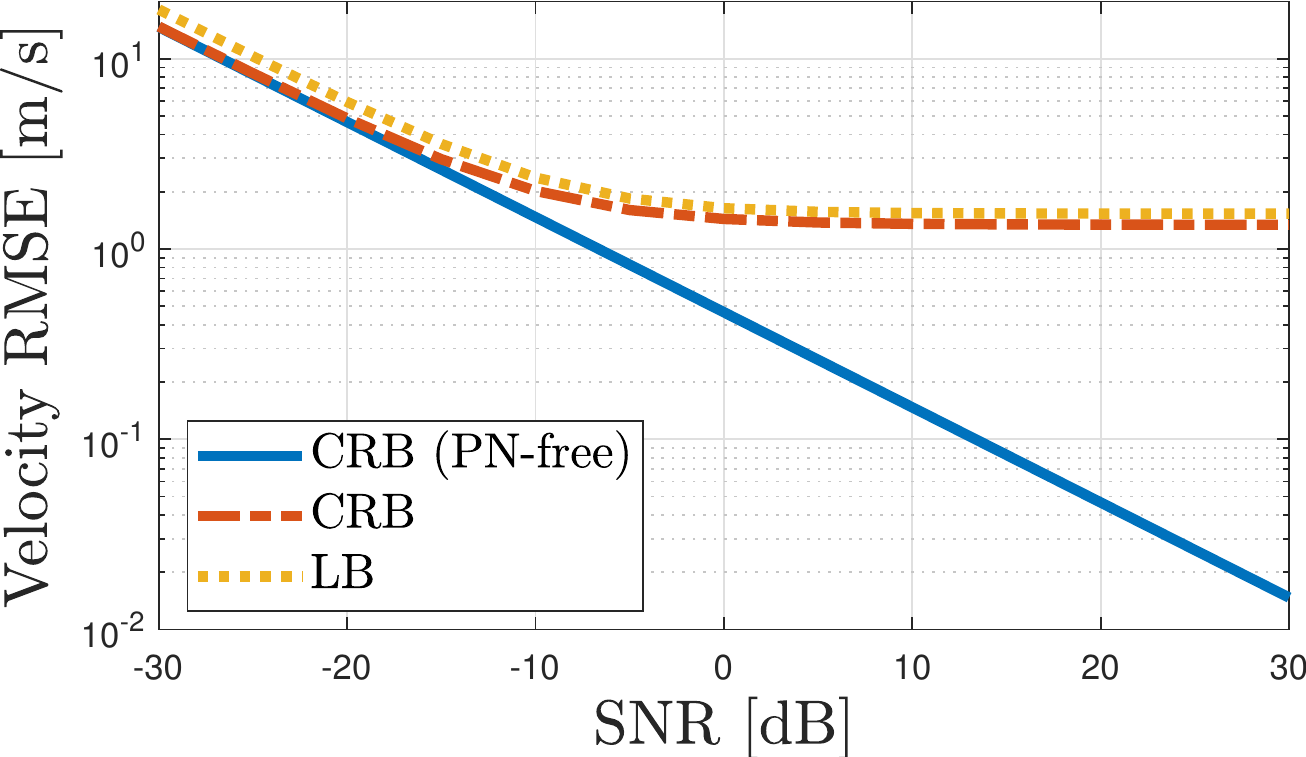}
		}
		\end{center}
		\vspace{-0.2in}
        \caption{Bounds on \subref{fig_range_rmse_r30_v20_f3dB_200_FRO} range and \subref{fig_vel_rmse_r30_v20_f3dB_200_FRO} velocity accuracy with respect to SNR for FRO with $\fdb = 100 \, \rm{kHz}$.} 
        \label{fig_rmse_r30_v20_f3dB_200_FRO}
        \vspace{-0.1in}
\end{figure}

\begin{figure}[t]
        \begin{center}
        \subfigure[]{
			 \label{fig_range_rmse_r30_v20_f3dB_200_PLL}
			 \includegraphics[width=0.38\textwidth]{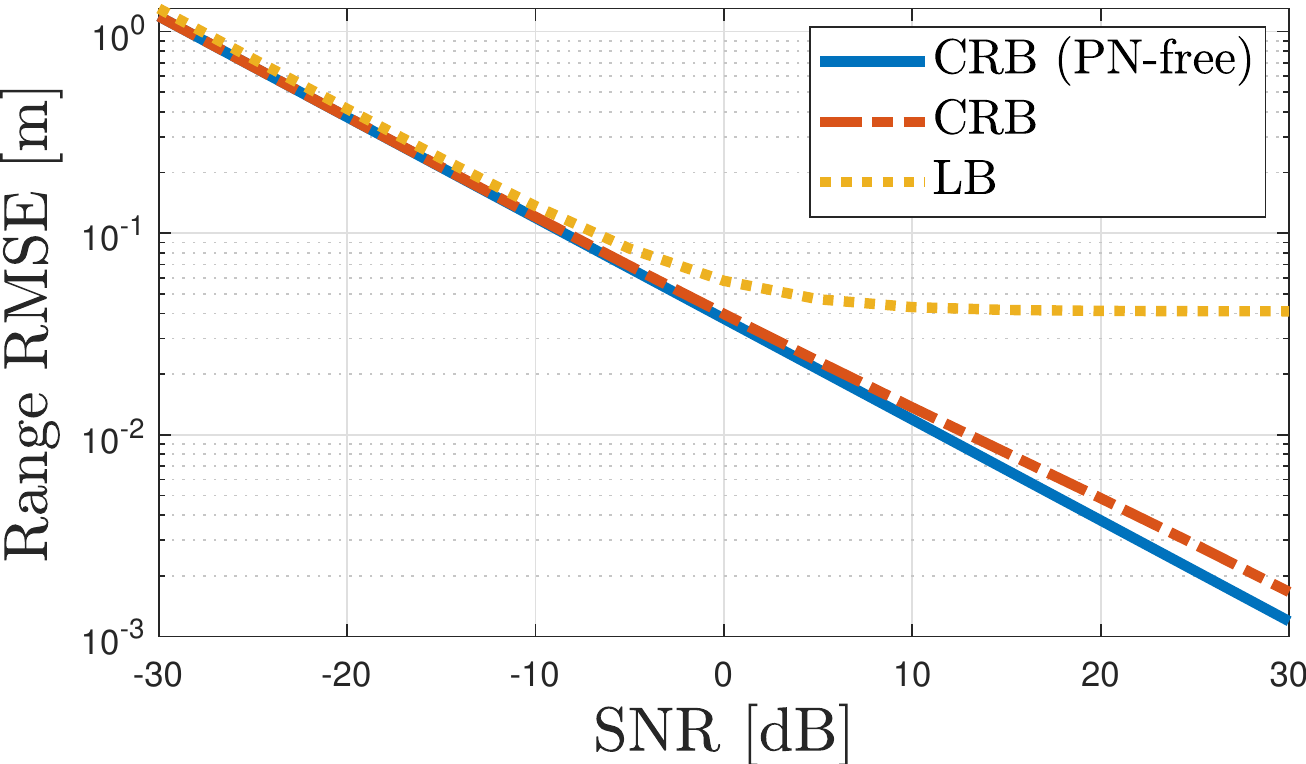}
		}
        \subfigure[]{
			 \label{fig_vel_rmse_r30_v20_f3dB_200_PLL}
			 \includegraphics[width=0.38\textwidth]{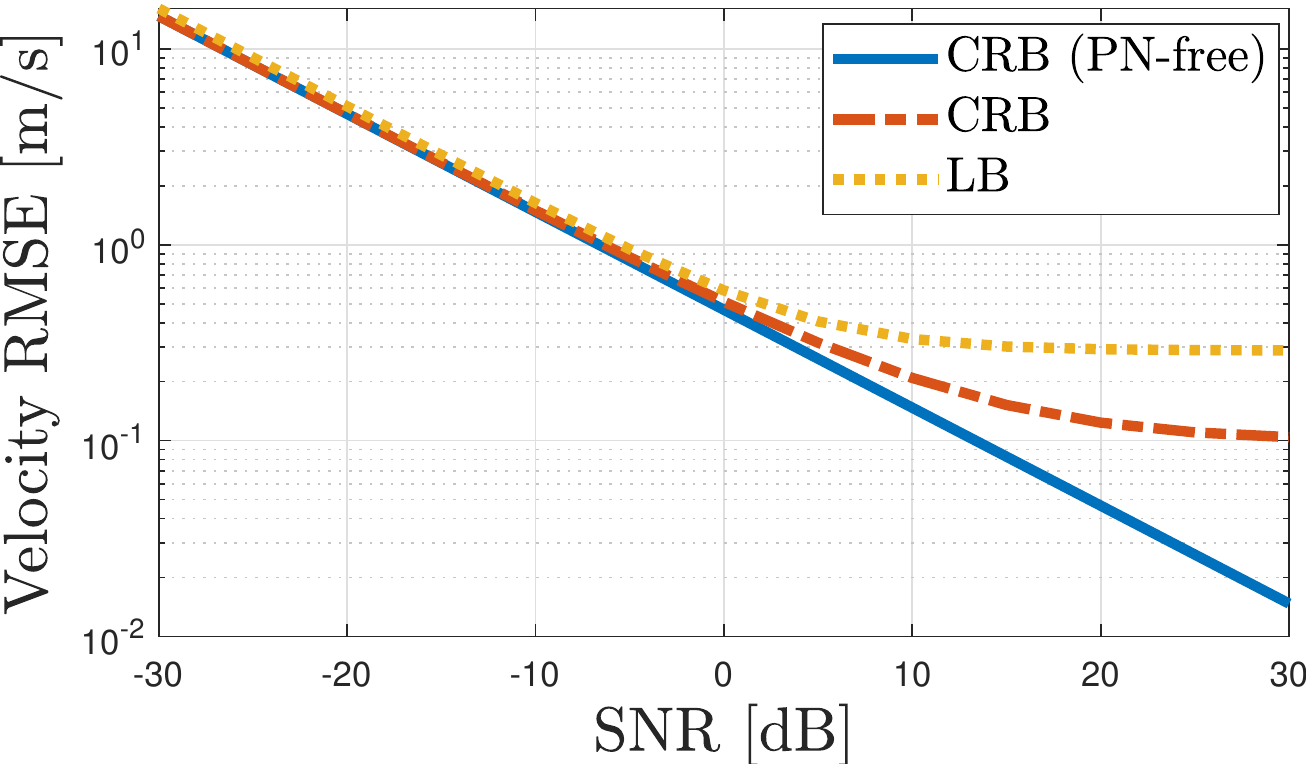}
		}
		\end{center}
		\vspace{-0.2in}
        \caption{Bounds on \subref{fig_range_rmse_r30_v20_f3dB_200_FRO} range and \subref{fig_vel_rmse_r30_v20_f3dB_200_FRO} velocity accuracy with respect to SNR for PLL with $\fdb = 100 \, \rm{kHz}$ and $\floop = 1 \, \rm{MHz}$.} 
        \label{fig_rmse_r30_v20_f3dB_200_PLL}
        \vspace{-0.1in}
\end{figure}

In Fig.~\ref{fig_rmse_r30_v20_f3dB_200_FRO} and Fig.~\ref{fig_rmse_r30_v20_f3dB_200_PLL}, we show the range and velocity estimation performances\footnote{The y-axes are labeled as root-mean-squared error (RMSE) to have a common name for all the bounds.} with respect to SNR for FRO and PLL architectures, respectively. Comparing the CRB (PN-free) and the CRB curves, it is observed that when the receiver is aware of the existence of PN, range accuracy is only slightly degraded, whereas velocity accuracy reaches a plateau at a certain SNR value, leading to considerable losses in performance at high SNRs. This results from high (low) correlation of PN in fast-time (slow-time) \cite{ofdm_pn_2022}. Therefore, the impact of PN on ranging accuracy can be mitigated by utilizing the PN statistics in the estimation process. 

Looking at the LB curves, we see that being unaware of the existence of PN causes performance saturation in both range and velocity estimation at medium and high SNR regimes. This is due to the model mismatch between the true and assumed models in Sec.~\ref{sec_true_assumed}, leading to the SNR-independent bias term in \eqref{eq_lb}, which dominates the LB expression as the SNR increases \cite{ozturk2022ris}. Hence, PN can be safely ignored at low SNRs, while it should be compensated for in radar processing in medium-to-high SNR regimes, especially for emerging 6G sensing applications with stringent accuracy requirements \cite{6G_HexaX_Access}. Finally, a comparative analysis of the CRB and LB curves indicates an important distinction between FRO and PLL. Namely, in velocity estimation, a certain degree of performance loss due to PN can be recovered in the case of PLL, whereas with FRO almost no improvement over the PN-unaware case (i.e., LB) is possible by exploiting the PN statistics. The reason is that PN in FRO has no slow-time correlation to exploit, unlike in PLL (see \eqref{eq_fro_pll_variance} and \cite[Lemma~2]{ofdm_pn_2022} for details).

\subsection{Performance with respect to Target Range}
\begin{figure}[t]
        \begin{center}
        \subfigure[]{
			 \label{fig_range_range_FRO}
			 \includegraphics[width=0.38\textwidth]{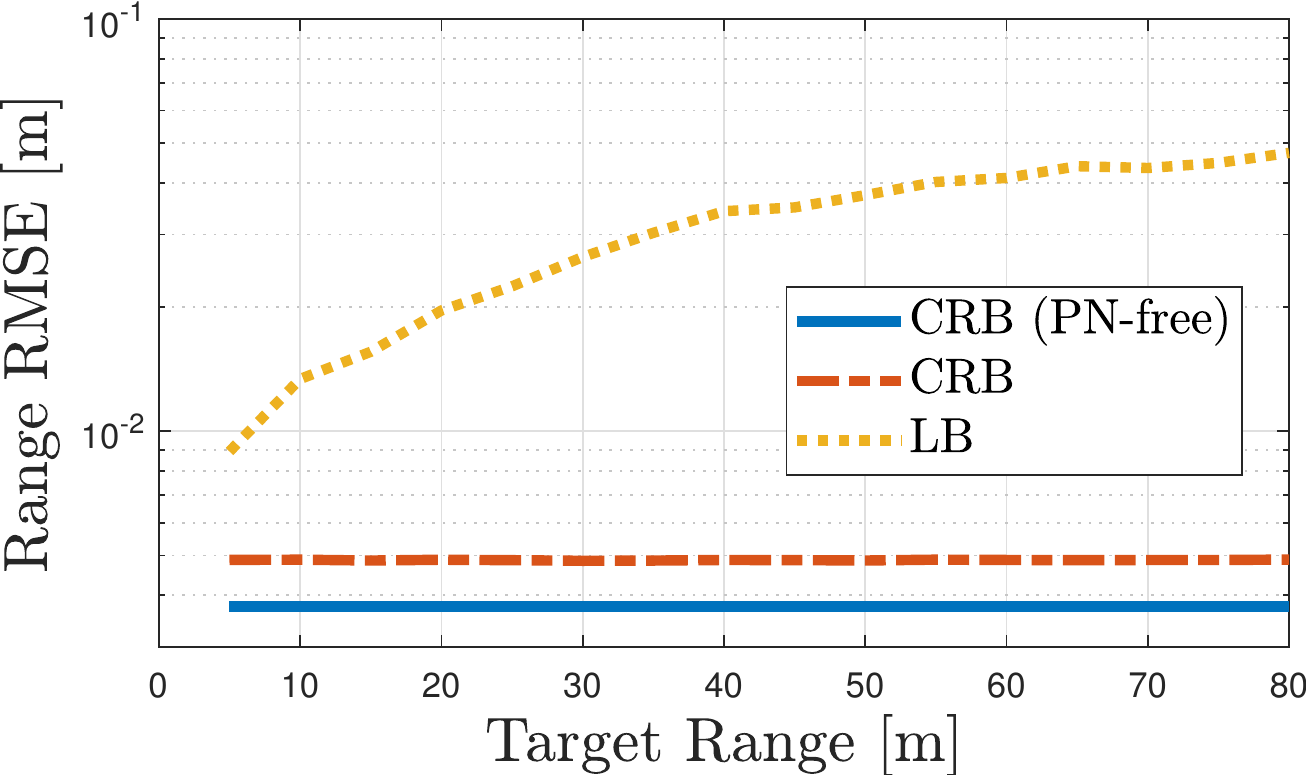}
		}
        \subfigure[]{
			 \label{fig_range_vel_FRO}
			 \includegraphics[width=0.38\textwidth]{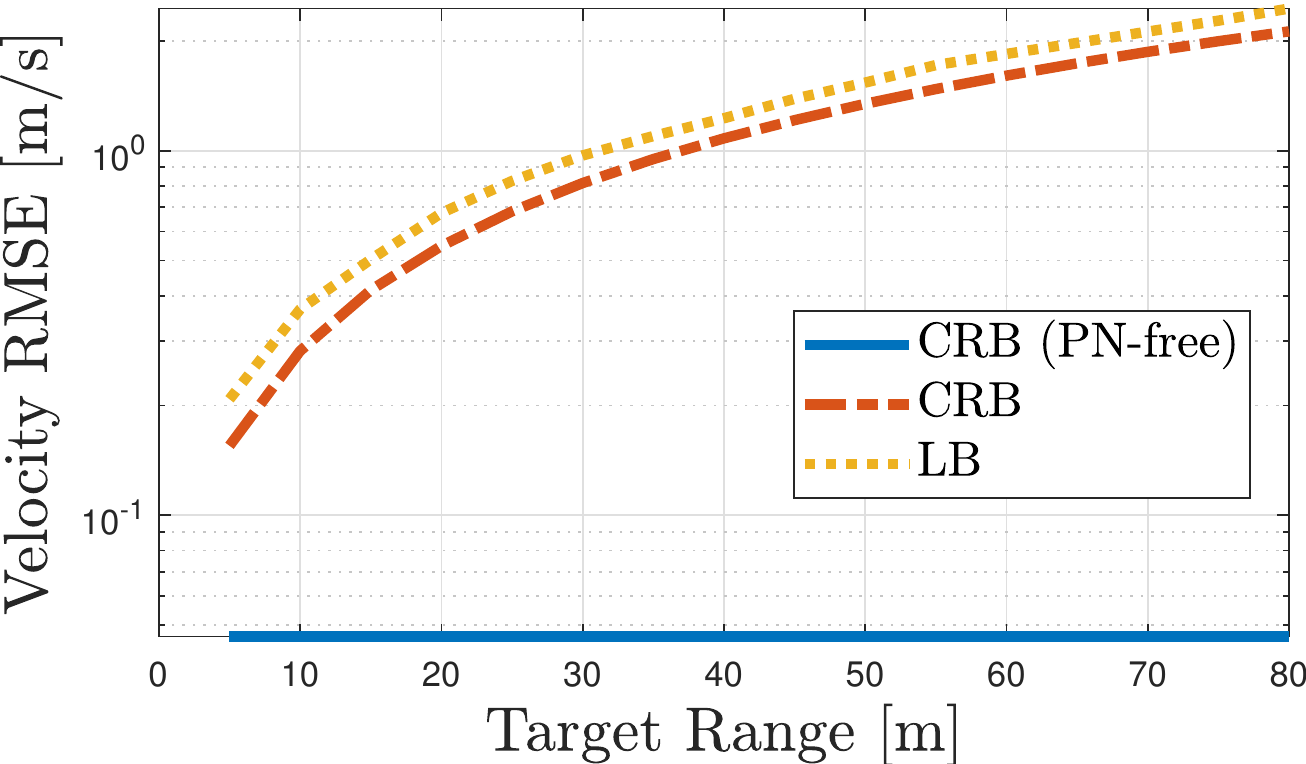}
		}
		\end{center}
		\vspace{-0.2in}
        \caption{Bounds on \subref{fig_range_range_FRO} range and \subref{fig_range_vel_FRO} velocity accuracy with respect to target range at $\snr = 20 \, \rm{dB}$ for FRO with $\fdb = 100 \, \rm{kHz}$.} 
        \label{fig_range_FRO_all}
        \vspace{-0.1in}
\end{figure}

In this part, considering the delay-dependent PN statistics given in \eqref{eq_bxi_stat}, we evaluate the radar performance under PN with respect to target range for fixed SNR, as illustrated in Fig.~\ref{fig_range_FRO_all} for FRO with $\fdb = 100 \, \rm{kHz}$. For the LB, we observe deteriorating performance in both range and velocity estimation as the target moves away from the radar due to increasing PN variance with range, as seen from \eqref{eq_fro_pll_variance}. For the CRB, ranging accuracy stays constant with target range because PN can be compensated effectively in fast-time through its high level of correlation. On the other hand, the CRB on velocity increases with target range due to the absence of slow-time PN correlation for FRO.


\subsection{Performance with respect to Oscillator Quality}
\begin{figure}[t]
        \begin{center}
        \subfigure[]{
			 \label{fig_range_f3db_FRO}
			 \includegraphics[width=0.38\textwidth]{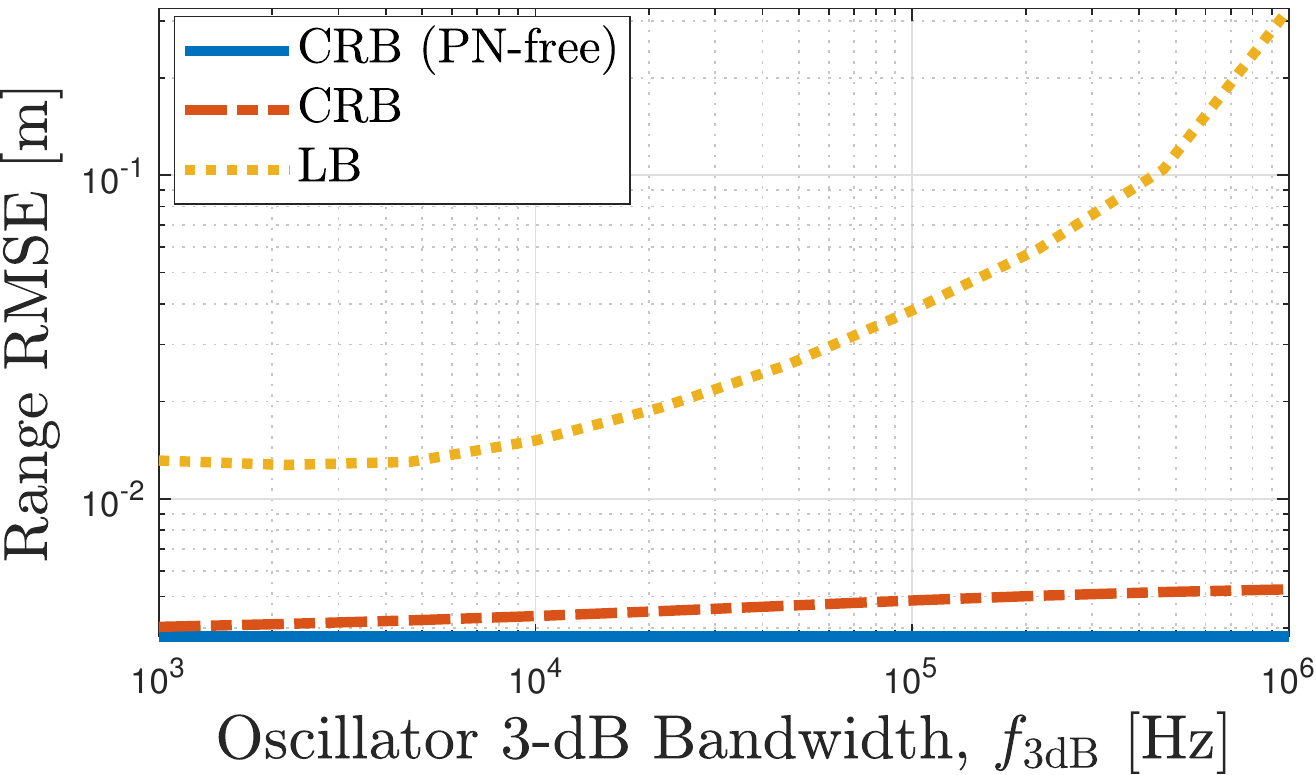}
		}
        \subfigure[]{
			 \label{fig_vel_f3db_FRO}
			 \includegraphics[width=0.38\textwidth]{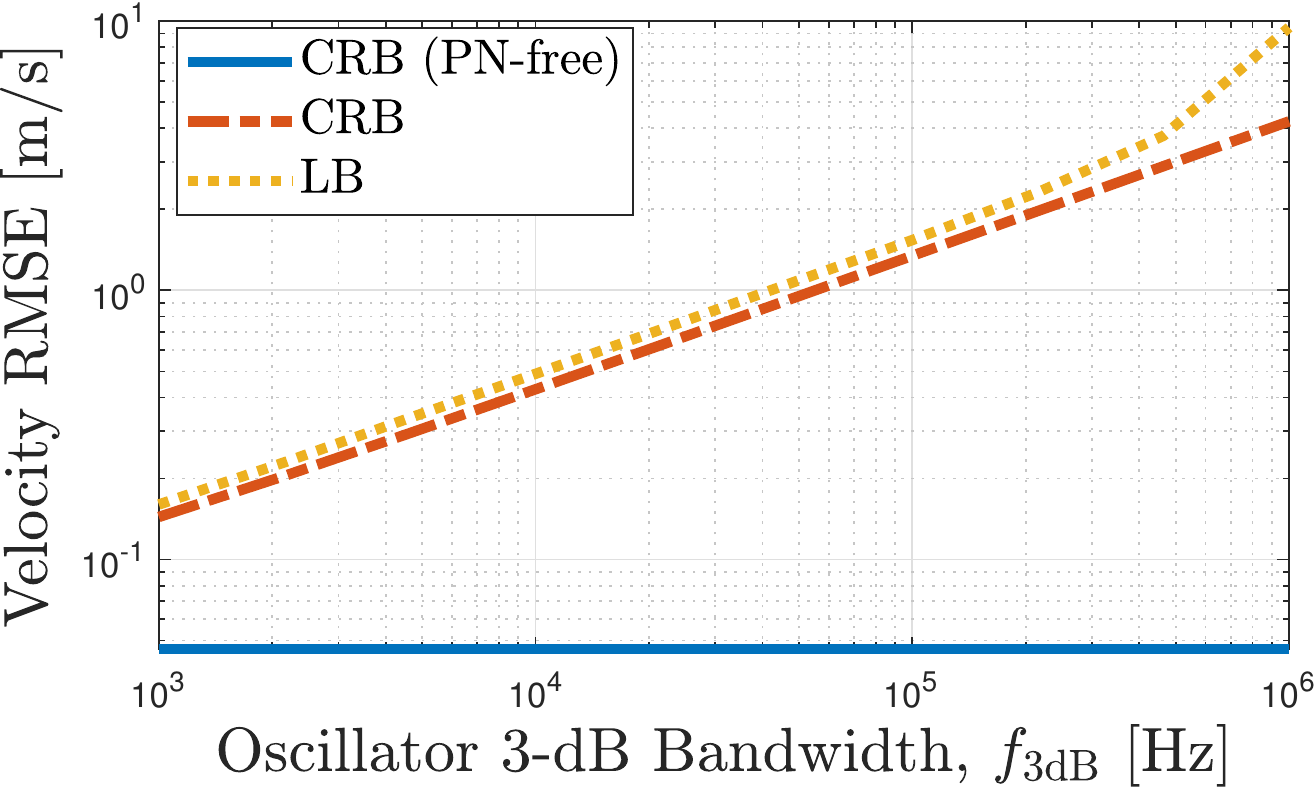}
		}
		\end{center}
		\vspace{-0.2in}
        \caption{Bounds on \subref{fig_range_floop_PLL} range and \subref{fig_vel_floop_PLL} velocity accuracy with respect to $\fdb$ of FRO at $\snr = 20 \, \rm{dB}$.} 
        \label{fig_range_vel_f3db_FRO}
        \vspace{-0.12in}
\end{figure}

\begin{figure}[t]
        \begin{center}
        \subfigure[]{
			 \label{fig_range_floop_PLL}
			 \includegraphics[width=0.38\textwidth]{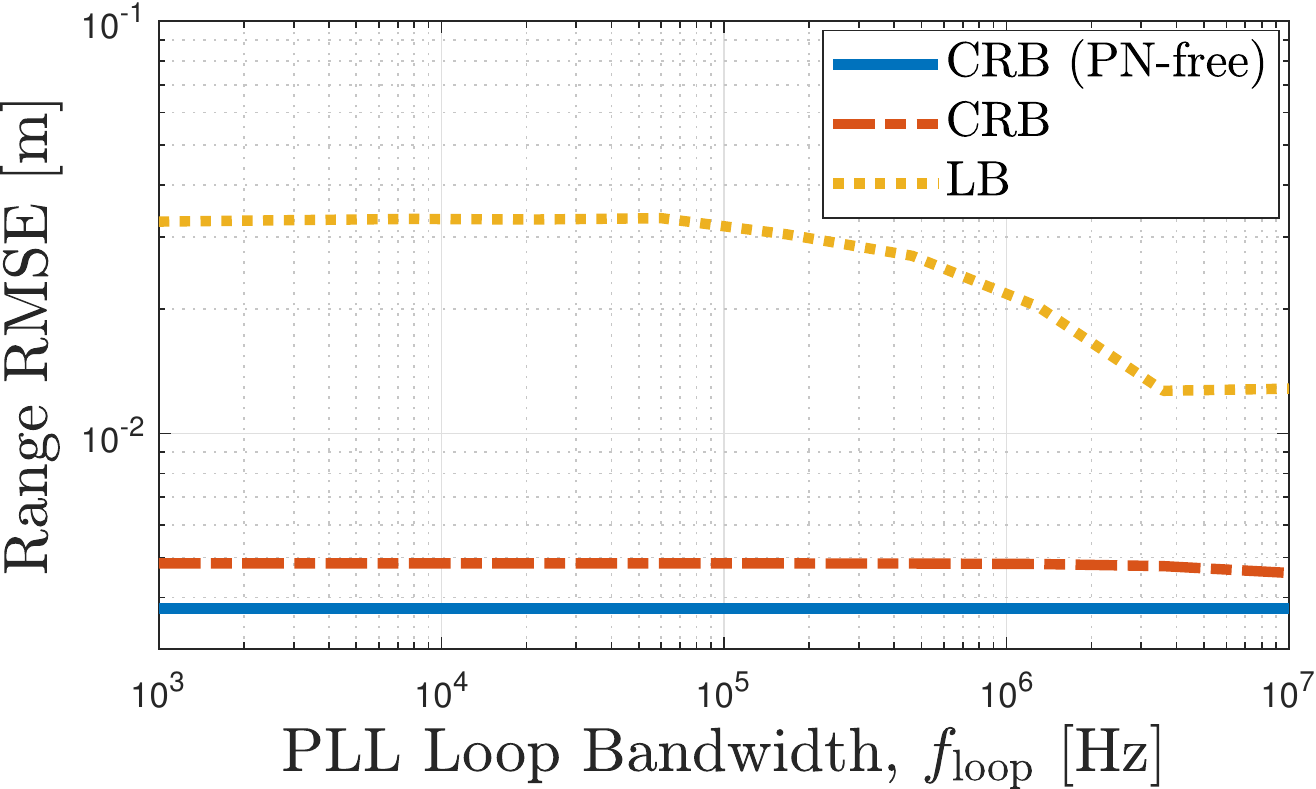}
		}
        \subfigure[]{
			 \label{fig_vel_floop_PLL}
			 \includegraphics[width=0.38\textwidth]{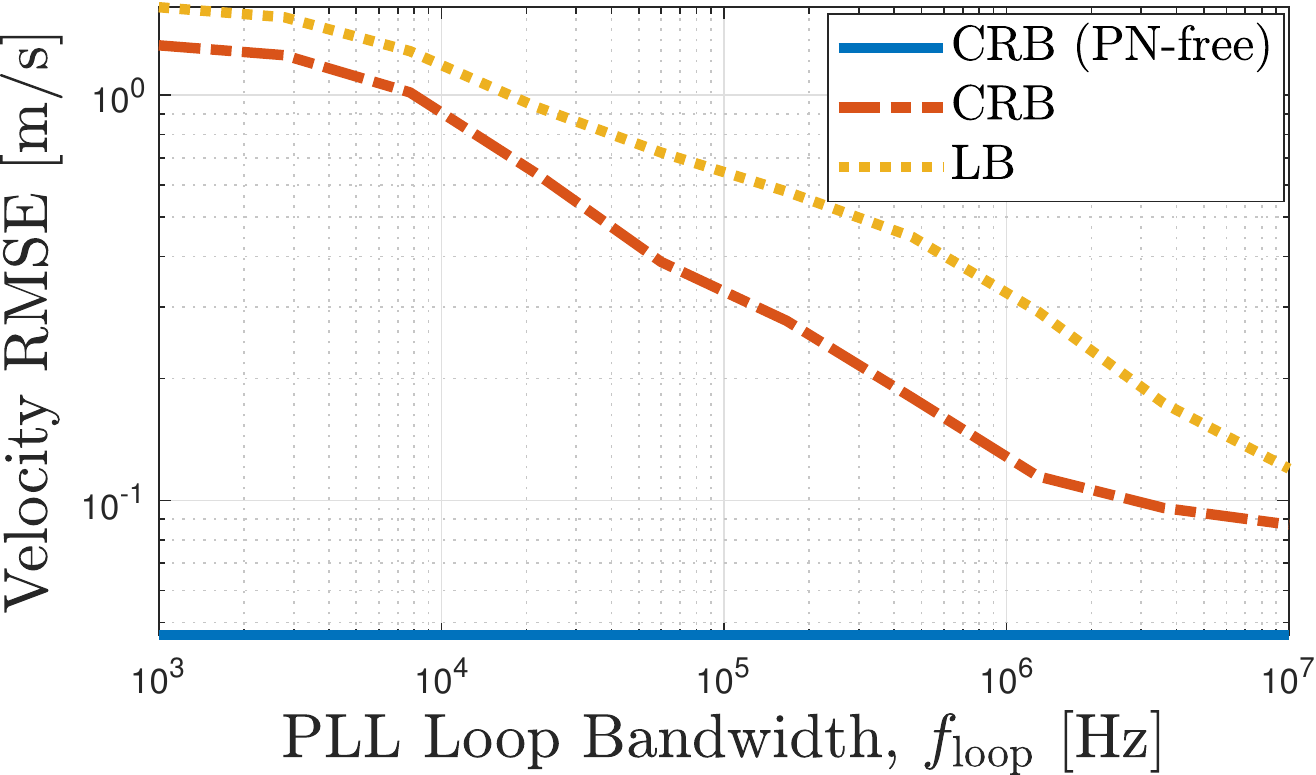}
		}
		\end{center}
		\vspace{-0.2in}
        \caption{Bounds on \subref{fig_range_floop_PLL} range and \subref{fig_vel_floop_PLL} velocity accuracy with respect to $\floop$ of PLL with $\fdb = 100 \, \rm{kHz}$ at $\snr = 20 \, \rm{dB}$.} 
        \label{fig_range_vel_floop_PLL}
        \vspace{-0.1in}
\end{figure}

In this part, we assess the impact of PN for different levels of oscillator quality. Fig.~\ref{fig_range_vel_f3db_FRO} and Fig.~\ref{fig_range_vel_floop_PLL} show the theoretical limits on range and velocity estimation against $\fdb$ and $\floop$ for FRO and PLL, respectively. It is seen from the CRB curves that ranging accuracy remains almost constant with respect to worsening oscillator quality (i.e., increasing $\fdb$ and decreasing $\floop$), indicating that PN can be mitigated in fast-time by using the PN statistics. However, due to similar reasons as in the previous subsections, the CRB on velocity increases with decreasing oscillator quality. Moreover, in the case of the LB, the range and velocity accuracy exhibit similar trends due to model misspecification, as expected.

\section{Concluding Remarks}\label{sec_conc}
We have investigated the impact of PN on range and velocity accuracy in an OFDM ISAC system by leveraging the CRB and MCRB tools. It has been observed that when PN is ignored in radar processing (i.e., model mismatch), it leads to an error floor in both range and velocity estimation, which can create a bottleneck for high-accuracy ($\rm{cm}$-level or $\rm{dm/s}$-level) sensing applications towards dual-functional 6G networks. When PN statistics are taken into account in the estimation process, the radar receiver experiences only marginal degradation in ranging accuracy due to high correlation of PN in fast-time, as opposed to substantial loss in velocity accuracy, especially at high SNRs. Moreover, the range correlation effect in monostatic sensing has been observed through delay-dependent PN statistics, which leads to lower accuracy for farther targets.

\section*{Acknowledgments}
This work is supported, in part, by Vinnova Grant 2021-02568 and MSCA-IF grant 888913 (OTFS-RADCOM). 

\bibliographystyle{IEEEtran}
\bibliography{ofdm_dfrc}

\begin{thebibliography}{10}
\providecommand{\url}[1]{#1}
\csname url@samestyle\endcsname
\providecommand{\newblock}{\relax}
\providecommand{\bibinfo}[2]{#2}
\providecommand{\BIBentrySTDinterwordspacing}{\spaceskip=0pt\relax}
\providecommand{\BIBentryALTinterwordstretchfactor}{4}
\providecommand{\BIBentryALTinterwordspacing}{\spaceskip=\fontdimen2\font plus
\BIBentryALTinterwordstretchfactor\fontdimen3\font minus
  \fontdimen4\font\relax}
\providecommand{\BIBforeignlanguage}[2]{{%
\expandafter\ifx\csname l@#1\endcsname\relax
\typeout{** WARNING: IEEEtran.bst: No hyphenation pattern has been}%
\typeout{** loaded for the language `#1'. Using the pattern for}%
\typeout{** the default language instead.}%
\else
\language=\csname l@#1\endcsname
\fi
#2}}
\providecommand{\BIBdecl}{\relax}
\BIBdecl

\bibitem{Fan_ISAC_6G_JSAC_2022}
F.~Liu \emph{et~al.}, ``Integrated sensing and communications: Toward
  dual-functional wireless networks for {6G} and beyond,'' \emph{IEEE Journal
  on Selected Areas in Communications}, vol.~40, no.~6, pp. 1728--1767, 2022.

\bibitem{JCAS_Survey_2022}
J.~A. Zhang \emph{et~al.}, ``Enabling joint communication and radar sensing in
  mobile networks—a survey,'' \emph{IEEE Communications Surveys Tutorials},
  vol.~24, no.~1, pp. 306--345, 2022.

\bibitem{hexax_pimrc_2021}
H.~Wymeersch \emph{et~al.}, ``Integration of communication and sensing in {6G}:
  a joint industrial and academic perspective,'' in \emph{2021 IEEE 32nd Annual
  International Symposium on Personal, Indoor and Mobile Radio Communications
  (PIMRC)}, 2021, pp. 1--7.

\bibitem{RF_JCS_2021}
F.~Bozorgi \emph{et~al.}, ``{RF} front-end challenges for joint communication
  and radar sensing,'' in \emph{1st IEEE Int. Online Symp. Joint Commun.
  Sens.}, Feb. 2021.

\bibitem{ofdm_pn_2022}
M.~F. Keskin \emph{et~al.}, ``Monostatic sensing with {OFDM} under phase noise:
  From mitigation to exploitation,'' \emph{arXiv preprint arXiv:2205.08376v2},
  2022.

\bibitem{PN_FMCW_2019}
K.~Siddiq \emph{et~al.}, ``Phase noise in {FMCW} radar systems,'' \emph{IEEE
  Transactions on Aerospace and Electronic Systems}, vol.~55, no.~1, pp.
  70--81, 2019.

\bibitem{Canan_SPM_2020}
C.~{Aydogdu} \emph{et~al.}, ``Radar interference mitigation for automated
  driving: Exploring proactive strategies,'' \emph{IEEE Signal Processing
  Magazine}, vol.~37, no.~4, pp. 72--84, 2020.

\bibitem{PN_OFDM_Sayed_TSP_2007}
Q.~Zou \emph{et~al.}, ``Compensation of phase noise in {OFDM} wireless
  systems,'' \emph{IEEE Transactions on Signal Processing}, vol.~55, no.~11,
  pp. 5407--5424, 2007.

\bibitem{VI_PN_TSP_2007}
D.~D. Lin \emph{et~al.}, ``The variational inference approach to joint data
  detection and phase noise estimation in {OFDM},'' \emph{IEEE Transactions on
  Signal Processing}, vol.~55, no.~5, pp. 1862--1874, 2007.

\bibitem{PN_Spectral_ICI_TSP_2010}
P.~Rabiei \emph{et~al.}, ``A non-iterative technique for phase noise {ICI}
  mitigation in packet-based {OFDM} systems,'' \emph{IEEE Transactions on
  Signal Processing}, vol.~58, no.~11, pp. 5945--5950, 2010.

\bibitem{PN_OFDM_TSP_2017}
A.~Leshem \emph{et~al.}, ``Phase noise compensation for {OFDM} systems,''
  \emph{IEEE Transactions on Signal Processing}, vol.~65, no.~21, pp.
  5675--5686, 2017.

\bibitem{OFDM_JRC_PN_JLT_2022}
Z.~Xue \emph{et~al.}, ``{OFDM} radar and communication joint system using
  opto-electronic oscillator with phase noise degradation analysis and
  mitigation,'' \emph{Journal of Lightwave Technology}, pp. 1--1, 2022.

\bibitem{SC_OFDM_PN}
B.~Schweizer \emph{et~al.}, ``On hardware implementations of stepped-carrier
  {OFDM} radars,'' in \emph{2018 IEEE/MTT-S International Microwave Symposium -
  IMS}, 2018, pp. 891--894.

\bibitem{Richmond_TSP_2015}
C.~D. Richmond \emph{et~al.}, ``Parameter bounds on estimation accuracy under
  model misspecification,'' \emph{IEEE Transactions on Signal Processing},
  vol.~63, no.~9, pp. 2263--2278, 2015.

\bibitem{Fortunati2017}
S.~Fortunati \emph{et~al.}, ``Performance bounds for parameter estimation under
  misspecified models: {F}undamental findings and applications,'' \emph{IEEE
  Signal Process. Mag.}, vol.~34, no.~6, pp. 142--157, 2017.

\bibitem{hybrid_CRB_LSP_2008}
S.~Bay \emph{et~al.}, ``On the hybrid {Cramér Rao} bound and its application
  to dynamical phase estimation,'' \emph{IEEE Signal Processing Letters},
  vol.~15, pp. 453--456, 2008.

\bibitem{Hybrid_ML_MAP_TSP}
Y.~Noam \emph{et~al.}, ``Notes on the tightness of the hybrid {Cramér–Rao}
  lower bound,'' \emph{IEEE Transactions on Signal Processing}, vol.~57, no.~6,
  pp. 2074--2084, 2009.

\bibitem{RadCom_Proc_IEEE_2011}
C.~{Sturm} \emph{et~al.}, ``Waveform design and signal processing aspects for
  fusion of wireless communications and radar sensing,'' \emph{Proceedings of
  the IEEE}, vol.~99, no.~7, pp. 1236--1259, July 2011.

\bibitem{PN_mmWave_OFDM_TWC_2022}
J.~Rodríguez-Fernández, ``Joint synchronization and compressive channel
  estimation for frequency-selective hybrid mmwave {MIMO} systems,'' \emph{IEEE
  Transactions on Wireless Communications}, vol.~21, no.~1, pp. 548--562, 2022.

\bibitem{PN_OFDM_PLL_TCOM_2007}
D.~Petrovic \emph{et~al.}, ``Effects of phase noise on {OFDM} systems with and
  without {PLL}: Characterization and compensation,'' \emph{IEEE Transactions
  on Communications}, vol.~55, no.~8, pp. 1607--1616, 2007.

\bibitem{OFDM_PN_HCRB_2014}
O.~H. Salim \emph{et~al.}, ``Channel, phase noise, and frequency offset in
  {OFDM} systems: Joint estimation, data detection, and hybrid cramér-rao
  lower bound,'' \emph{IEEE Transactions on Communications}, vol.~62, no.~9,
  pp. 3311--3325, 2014.

\bibitem{General_Multicarrier_Radar_TSP_2016}
M.~{Bică} \emph{et~al.}, ``Generalized multicarrier radar: Models and
  performance,'' \emph{IEEE Transactions on Signal Processing}, vol.~64,
  no.~17, pp. 4389--4402, Sep. 2016.

\bibitem{PN_2006}
A.~{Chorti} \emph{et~al.}, ``A spectral model for {RF} oscillators with
  power-law phase noise,'' \emph{IEEE Transactions on Circuits and Systems I:
  Regular Papers}, vol.~53, no.~9, pp. 1989--1999, Sep. 2006.

\bibitem{PN_SI_TSP_2017}
X.~Quan \emph{et~al.}, ``Impacts of phase noise on digital self-interference
  cancellation in full-duplex communications,'' \emph{IEEE Transactions on
  Signal Processing}, vol.~65, no.~7, pp. 1881--1893, 2017.

\bibitem{SPM_PN_2019}
M.~Gerstmair \emph{et~al.}, ``On the safe road toward autonomous driving: Phase
  noise monitoring in radar sensors for functional safety compliance,''
  \emph{IEEE Signal Processing Magazine}, vol.~36, no.~5, pp. 60--70, 2019.

\bibitem{Firat_OFDM_2012}
R.~F. {Tigrek} \emph{et~al.}, ``{OFDM} signals as the radar waveform to solve
  {Doppler} ambiguity,'' \emph{IEEE Transactions on Aerospace and Electronic
  Systems}, vol.~48, no.~1, pp. 130--143, Jan 2012.

\bibitem{ICI_OFDM_TSP_2020}
F.~Zhang \emph{et~al.}, ``Joint range and velocity estimation with intrapulse
  and intersubcarrier {Doppler} effects for {OFDM}-based {RadCom} systems,''
  \emph{IEEE Transactions on Signal Processing}, vol.~68, pp. 662--675, 2020.

\bibitem{MIMO_OFDM_ICI_JSTSP_2021}
M.~F. Keskin \emph{et~al.}, ``{MIMO-OFDM} joint radar-communications: Is {ICI}
  friend or foe?'' \emph{IEEE Journal of Selected Topics in Signal Processing},
  vol.~15, no.~6, pp. 1393--1408, 2021.

\bibitem{OFDM_ICI_TVT_2017}
G.~Hakobyan \emph{et~al.}, ``A novel intercarrier-interference free signal
  processing scheme for {OFDM} radar,'' \emph{IEEE Transactions on Vehicular
  Technology}, vol.~67, no.~6, pp. 5158--5167, 2017.

\bibitem{OFDM_Radar_Phd_2014}
M.~Braun, ``{OFDM} radar algorithms in mobile communication networks,''
  \emph{Karlsruher Institutes f{\"u}r Technologie}, 2014.

\bibitem{SPM_JRC_2019}
K.~V. {Mishra} \emph{et~al.}, ``Toward millimeter-wave joint radar
  communications: A signal processing perspective,'' \emph{IEEE Signal
  Processing Magazine}, vol.~36, no.~5, pp. 100--114, Sep. 2019.

\bibitem{Passive_OFDM_2010}
C.~R. Berger \emph{et~al.}, ``Signal processing for passive radar using {OFDM}
  waveforms,'' \emph{IEEE Journal of Selected Topics in Signal Processing},
  vol.~4, no.~1, pp. 226--238, 2010.

\bibitem{OTFS_RadCom_TWC_2020}
L.~Gaudio \emph{et~al.}, ``On the effectiveness of {OTFS} for joint radar
  parameter estimation and communication,'' \emph{IEEE Transactions on Wireless
  Communications}, vol.~19, no.~9, pp. 5951--5965, 2020.

\bibitem{OFDM_DFRC_TSP_2021}
M.~F. Keskin \emph{et~al.}, ``Limited feedforward waveform design for {OFDM}
  dual-functional radar-communications,'' \emph{IEEE Transactions on Signal
  Processing}, vol.~69, pp. 2955--2970, 2021.

\bibitem{Demir_PN_2006}
A.~{Demir}, ``Computing timing jitter from phase noise spectra for oscillators
  and phase-locked loops with white and $1/f$ noise,'' \emph{IEEE Transactions
  on Circuits and Systems I: Regular Papers}, vol.~53, no.~9, pp. 1869--1884,
  Sep. 2006.

\bibitem{DPN_93}
A.~Murat \emph{et~al.}, ``Phase-noise-induced performance limits for {DPSK}
  modulation with and without frequency feedback,'' \emph{Journal of Lightwave
  Technology}, vol.~11, no.~2, pp. 290--302, 1993.

\bibitem{kay1993fundamentals}
S.~M. Kay, \emph{Fundamentals of Statistical Signal Processing: Estimation
  Theory}.\hskip 1em plus 0.5em minus 0.4em\relax Prentice Hall, 1993.

\bibitem{Fortunati2018Chapter4}
S.~Fortunati \emph{et~al.}, ``{Chapter 4: Parameter bounds under misspecified
  models for adaptive radar detection},'' in \emph{Academic Press Library in
  Signal Processing, Volume 7}, R.~Chellappa \emph{et~al.}, Eds.\hskip 1em plus
  0.5em minus 0.4em\relax Academic Press, 2018, pp. 197--252.

\bibitem{MCRB_delay_ICASSP_2020}
F.~Roemer, ``Misspecified {Cramer-Rao} bound for delay estimation with a
  mismatched waveform: {A} case study,'' in \emph{IEEE Int. Conf. Acoustics,
  Speech Signal Process.}, 2020, pp. 5994--5998.

\bibitem{ozturk2022ris}
C.~Ozturk \emph{et~al.}, ``{RIS}-aided near-field localization under
  phase-dependent amplitude variations,'' \emph{arXiv preprint
  arXiv:2204.12783}, 2022.

\bibitem{TR_38211}
3GPP, ``{NR} physical channels and modulation,'' 3GPP TR38.211 V16.1.0, Sophia
  Antipolis, France, Tech. Rep., 2020.

\bibitem{6G_HexaX_Access}
M.~A. Uusitalo \emph{et~al.}, ``{6G} vision, value, use cases and technologies
  from european {6G} flagship project {Hexa-X},'' \emph{IEEE Access}, vol.~9,
  pp. 160\,004--160\,020, 2021.

\end{thebibliography}

\end{document}